# Topological Phase Transition in Mechanical Honeycomb Lattice

Yi Chen, Xiaoning Liu* and Gengkai Hu*

Key Laboratory of Dynamics and Control of Flight Vehicle, Ministry of Education, School of Aerospace Engineering, Beijing Institute of Technology, Beijing 100081, China

**Abstract**

Topological materials provide a new tool to direct wave energy with unprecedented precision and robustness. Three elastic topological phases, the valley Hall, Chern and spin Hall insulators, are currently studied, and they are achieved separately in rather distinct configurations. Here, we explore analytically various topological phase transitions for in-plane elastic wave in a unified mass-spring honeycomb lattice. It is demonstrated that the three elastic topological phases can be realized in this single lattice by designing mass, stiffness or introducing Coriolis' effect. In particular, the interface between valley Hall and Chern insulators is found to support topological interface mode for the first time. Perturbation method is used to derive the analytic effective continuum model in the neighbor of band degeneracy, and the physics in topological phase transitions are revealed through evaluation of topological invariants. The topologically protected interface states, their decaying profile as well as the pseudo-spin-indicating polarization specific for elastic wave are systematically analyzed, and these results are further confirmed numerically by Bloch wave analysis of domain wall strip and transient simulation of finite sized sample. This study offers a concise and unified analytical model to explore topology nature of elastic wave, and can provide intuitive guidance to design of continuum mechanical topological materials.

* Corresponding authors: liuxn@bit.edu.cn, hugeng@bit.edu.cn

## 1. Introduction

During the past decades, there have been continuously active investigations for wave propagation in periodic materials, by exploring, for example, interplay between wave and microstructure through Bragg scattering, metasurfaces or metamaterials. Many exotic functions and devices for wave control have been conceived such as abnormal refraction, lensing or even cloaking. The mechanism of all these themes is engineering of band structure with deliberate microstructure design. Recently, the advent of topological phase, originally introduced in quantum systems of condensed matter (Haldane, 1988; Klitzing, et al., 1980; Thouless, et al., 1982), has attracted scientists to exploit topological property of band structure of periodic material. Topological phase is indexed by topological invariants, which are intrinsically linked to the geometrical property of bulk band characterized by Berry curvature (Kane and Mele, 2005a; Simon, 1983; Zhang, et al., 2013). A common feature is that domain walls between materials with distinct topological indices support edge waves, which are topologically protected and immune to backscattering of defects and sharp corners. This concept inspired from quantum systems is quickly extended to the realm of classical waves, such as light (Lu, et al., 2014), acoustic (He, et al., 2016; Lu, et al., 2016) and mechanical system (Prodan and Prodan, 2009; Nash et al., 2015; Süsstrunk and Huber, 2015), and even to zero-frequency floppy mode and targeted buckling in lattices with nontrivial topological polarization (Kane and Lubensky, 2014; Paulose et al., 2015). Topological mechanics is expected as a new paradigm for the design of wave devices robust to defects or disturbance (Huber, 2016).

Observation of topological phase in acoustic and mechanical systems is attractive not only because of its promising perspectives such as high-efficiency wave guiding, extraordinary information carrier (Jiang et al., 2018), but also because it constitutes a platform to explore similar phenomena in quantum physics without challenging atomic-scale techniques. One route to build topological phase is to introduce external source into the system in order to break the time reversal (TR) symmetry, thus mimicking the integer quantum Hall effect where the topological invariant is the Chern number. Prodan and Prodan (2009) draw the first mechanical analogue to the quantum Hall effect, and theoretically predicted robust edge state in a biological mechanical model with magnetic force to break the TR symmetry. The TR symmetry in acoustic fluids can be broken by rotating fans acting as flow-induced bias (Khanikaev, et al., 2015), while TR symmetry in mechanical lattice is usually broken by rotating gyroscope at each site (Wang et al., 2015a; Nash et al., 2015). A similar but much direct scheme is to place the whole mechanical lattice on a non-inertial rotating platform and exploits the Coriolis' force to break the TR symmetry (Kariyado and Hatsugai, 2015; Wang, et al., 2015b). Mechanical model with space-time modulated elastic constants was also demonstrated to be able to realize non-reciprocal wave propagation (Swinteck, et al., 2015).

Another route of creating topological phase requires however the preservation of TR symmetry, and shares the same principle of spin-orbit interaction with quantum spin Hall effect (QSHE) whose topological index is $Z_2$ invariant (Bernevig, et al., 2006; Kane and Mele, 2005a; b). This category of topological phase requires only passive element, however, for classical systems the design is subtle since spin is not an intrinsic degree of freedom (DOF). Süsstrunk and Huber (2015) obtained the topologically equivalent equation of dynamics of the Hofstadter lattice through unitary transformation, and they experimentally realized it by a double pendulum array with sophisticated connections. In a similar scheme, Pal et al. (2016) produced helical edge modes in a bi-layer lattice of rotational DOF via inter-layer coupling. Mousavi et al. (2015) numerically realized a spin Hall insulator for Lamb wave in a plate with perforations of distinct scales. Based on deterministic Dirac degeneracy in lattice with $C_{6v}$ symmetry, Wu and Hu (2015) proposed a supercell method to design the double fold Dirac cone and pseudo spin states in photonic system, which is also valid for acoustic system. A similar method is used in the design of an acoustic spin Hall insulator (He, et al., 2016). Besides, passive topological phase and edge mode can also be formed via the locally nontrivial topology of band structure in analogy with the quantum valley Hall effect (QVHE), in which an extra DOF, i.e. valley pseudo spin characterized by valley Chern number (Xiao, et al., 2007) can be pursued to generate topologically protected valley edge modes. QVHE is beneficial for reducing model complexity but is less robust owning to the accident inter-valley scattering. In this way, Lu et al. (2016) proposed a hexagonal sonic crystal with triangle scatters and experimentally demonstrated the valley transport, while the transverse motion of mechanical honeycomb lattice was studied by Pal and Ruzzene (2017), and experimentally validated by mass stubbed plate (Vila, et al., 2017). Recently, Chen et al. (2018) analytically showed that QVHE is also possible through designing spring stiffness in a Kagome lattice.

Though a variety of topological phases have been realized for mechanical systems, the proposed models are rather dispersed and much few works are related to the in-plane wave in elastic media. In this paper, we will show that the all known classes of topological phases can be realized for in-plane elastic wave in a unifying platform of mass-spring honeycomb lattice, which could be considered as a mechanical graphene (Neto, et al., 2009). In particular, we will show that the QVHE can be achieved by unequal masses on the lattice sites, and specifically, with the help of Coriolis' effect the switching and interplay between QVHE and Chern insulator can be explored for the first time. On the other hand, the nontrivial QSHE phase can be achieved in the same lattice by patterning the spring constants to manipulate the supercell and twofold Dirac degeneracy. Our primary goal is to advance the topological mechanics of in-plane elastic wave by analytical derivation of band structure, topological invariants, decaying profile of edge states and unique wave polarization. The proposed discrete model can provide useful guideline for the design of continuous topological elastic materials. A continuum version can be obtained by replacing the springs and masses with slender beams and circular disks, respectively. The beams should have a large slenderness ratio to ensure a small bending rigidity, thus can mimic a spring without transverse stiffness.

The paper is organized as follows: In Section 2, basic parameters of a discrete honeycomb lattice is introduced and a $k \cdot p$ perturbative effective model is derived. Section 3 is concerned with the design of QVHE insulator and Chern insulator through mass parameters and Coriolis' effect, while Section 4 is mainly concerned with designing elastic QSHE insulator through the control of stiffness in the lattice. The main findings are finally summarized in Section 5.

## 2. Mechanical honeycomb lattice and $k \cdot p$ perturbative effective model

The studied honeycomb mass-spring lattice is illustrated in Fig. 1(a), where a hexagonal unit cell has two sites respectively denoted as $p$ and $q$ with mass $m_p$ and $m_q$, and the nearest neighboring sites are linked by springs with stiffness $t$. The length and stiffness of the spring are all assumed to be $L_0 = 1$ and $t = 1$ unless specifically emphasized. Lattice vectors with length $a = \sqrt{3}L_0$ are $\mathbf{a}_1 = (a_x, -a_y) = (3L_0/2, -\sqrt{3}L_0/2)$ and $\mathbf{a}_2 = (3L_0/2, \sqrt{3}L_0/2)$, respectively. A hexagonal unit cell is highlighted in Fig. 1(a), while the first Brillouin zone (BZ) and reciprocal lattice vectors $\mathbf{b}_1 = (2\pi/(3L_0), -2\pi/(\sqrt{3}L_0))$, $\mathbf{b}_2 = (2\pi/(3L_0), +2\pi/(\sqrt{3}L_0))$ are depicted in Fig. 1(b). For a plain lattice with $m_p = m_q$ and uniform $t$, Dirac degeneracy is protected at inequivalent BZ corners $\mathbf{K}'$ and $\mathbf{K}$ owning to the $C_{6v}$ symmetry (Lu et al., 2014).

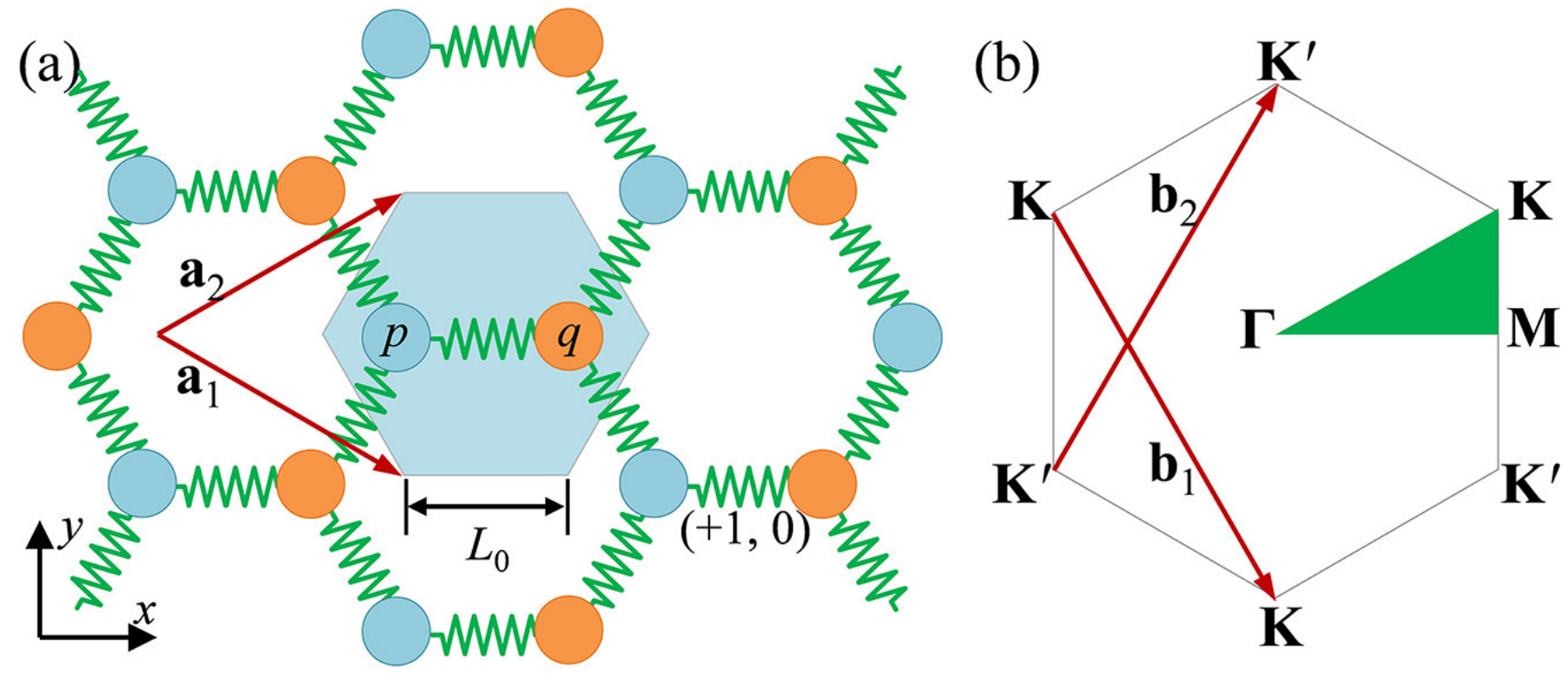

Figure 1: (a) Spring-mass honeycomb lattice with two sites in each unit cell; (b) First Brillouin zone with two inequivalent corners $\mathbf{K}'$ and $\mathbf{K} = -\mathbf{K}'$.

Here we consider in-plane elastic wave propagation in the lattice, and in-plane displacements DOF of mass $p$ and $q$ in the unit cell ($m$, $n$) are denoted as $\mathbf{u}^p_{m,n}$ and $\mathbf{u}^q_{m,n}$, respectively. The Lagrangian related to the reference cell (0,0) is,

$$L = \frac{1}{2} m_p (\dot{\mathbf{u}}^p_{m,n})^2 + \frac{1}{2} m_q (\dot{\mathbf{u}}^q_{m,n})^2 - \frac{1}{2} t \sum_{k,l,\alpha,\beta} |(\mathbf{u}^\alpha_{m,n} - \mathbf{u}^\beta_{k,l}) \cdot \mathbf{e}_t|^2 , \tag{1}$$

where, $\mathbf{e}_t$ stands for the director of spring linking two masses, and the summation runs over all cells adjacent to the reference one. Using the Lagrangian principle, the equation of motion of the reference cell can be obtained as,

$$\frac{d}{dt}\frac{\partial L}{\partial \dot{u}^\alpha_{m,n}} = \frac{\partial L}{\partial u^\alpha_{m,n}}, \quad \frac{d}{dt}\frac{\partial L}{\partial \dot{v}^\alpha_{m,n}} = \frac{\partial L}{\partial v^\alpha_{m,n}} \quad (\alpha = p,\ q) \tag{2}$$

We consider the Bloch wave solution of an infinite lattice in the form $\mathbf{u}_{m,n} = \mathbf{u} \exp(i\mathbf{k}\cdot(m\mathbf{a}_1 + n\mathbf{a}_2))$, where $\mathbf{k} = (k_x, k_y)$ is the Bloch wave vector and $\mathbf{u} = \{\mathbf{u}^p, \mathbf{u}^q\}^{\mathrm{T}}$ is the displacement vector of the reference unit cell (0,0). Time harmonic term $\exp(-i\omega t)$ is omitted as convention. Substituting the expression into Eq. (2) leads to the following eigenvalue problem

$$\left(\omega^2 \mathbf{M} + \mathbf{K}(\mathbf{k})\right)\mathbf{u} = 0 , \tag{3}$$

in which the 4 by 4 mass and stiffness matrices, $\mathbf{M}$ and $\mathbf{K}(\mathbf{k})$, are defined as

$$\mathbf{M} = \begin{pmatrix} m_p \mathbf{I} & 0 \\ 0 & m_q \mathbf{I} \end{pmatrix}, \quad \mathbf{K} = \begin{pmatrix} \mathbf{K}_{pp} & \mathbf{K}_{pq} \\ \mathbf{K}^\dagger_{pq} & \mathbf{K}_{qq} \end{pmatrix}, \tag{4}$$

$$\mathbf{K}_{pp} = \mathbf{K}_{qq} = -\frac{3t}{2}\mathbf{I}, \quad \mathbf{K}_{pq} = t\begin{pmatrix} 1 + \frac{1}{2}\exp(ik_x a_x)\cos k_y a_y & i\frac{\sqrt{3}}{2}\exp(ik_x a_x)\sin k_y a_y \\ i\frac{\sqrt{3}}{2}\exp(ik_x a_x)\sin k_y a_y & \frac{3}{2}\exp(ik_x a_x)\cos k_y a_y \end{pmatrix}. \tag{5}$$

Here dagger symbol represents complex conjugate transpose. Four branches of eigen-frequencies can be derived analytically,

$$\omega^2 = 0,\ \ \frac{3t}{4}\left(\frac{1}{m_p}+\frac{1}{m_q}\pm\frac{1}{m_p m_q}\sqrt{m_p^2+Ym_p m_q+m_q^2}\right),\ \ \frac{3t}{2}\left(\frac{1}{m_p}+\frac{1}{m_q}\right) \tag{6}$$

where $Y = 2/9(8\cos(k_x a)\cos(k_y a) + 4\cos(2k_y a) - 3)$. The eigenstates can also be obtained analytically, however are not presented here due to very cumbersome expression. The 1st and 4th bands are horizontal since each mass is connected to three springs hence the system has zero energy mode. For the 2nd and 3rd bands, extrema (valleys) are reached at **K** and **K′** points at BZ corner as $\omega = (3t/2m_p)^{1/2}$, $(3t/2m_q)^{1/2}$. At **K** valley, eigenstates corresponding to the two eigen-frequencies $\omega = (3t/2m_p)^{1/2}$ and $(3t/2m_q)^{1/2}$ are respectively,

$$\mathbf{u}_1(\mathbf{K}) = \frac{1}{\sqrt{2}}\{+i,1,0,0\}^{\mathrm{T}},\quad \mathbf{u}_2(\mathbf{K}) = \frac{1}{\sqrt{2}}\{0,0,-i,1\}^{\mathrm{T}}, \tag{7}$$

while eigenstates for **K'** valley are simply the conjugation as required by TR symmetry,

$$\mathbf{u}_1(\mathbf{K}') = \frac{1}{\sqrt{2}}\{-i,1,0,0\}^{\mathrm{T}},\quad \mathbf{u}_2(\mathbf{K}') = \frac{1}{\sqrt{2}}\{0,0,+i,1\}^{\mathrm{T}}. \tag{8}$$

The latter comes from the constraint of TR invariant Bloch Hamiltonian $TH(\mathbf{k}) = H(-\mathbf{k})T$, where $H(\mathbf{k}) = -\mathbf{M}^{-1}\mathbf{K}(\mathbf{k})$ from Eq. (3), and the time reversal operator $T$ is just the complex conjugate for the considered system without intrinsic spin DOFs (Asbóth et al., 2016). For the $\mathbf{u}_1(\mathbf{K})$ / $\mathbf{u}_1(\mathbf{K}')$ state, the site $p$ moves along a circular trajectory clockwise / counter-clockwise around the equilibrium position, while the site $q$ remains static. For the $\mathbf{u}_2(\mathbf{K})$ / $\mathbf{u}_2(\mathbf{K}')$ state, the site $q$ moves along a circular trajectory counter-clockwise / clockwise around the equilibrium position, while the site $p$ remains static. To understand the physics and topological phase transition adjacent to the valleys, $k \cdot p$ perturbation method used extensively in solid-state physics (Slonczewski and Weiss, 1958) will be employed in the following.

The second and third branches degenerate to a Dirac cone, at the **K** (**K'**) point, if the site masses $m_p = m_q$. If the two masses $m_p$ and $m_q$ are slightly different from each other $0 < |m_p - m_q| << 1$, the degeneracy will be lifted and a bandgap will be formed. The system can be considered as one perturbed from a degenerated system with equal mass $(m_p + m_q)/2$. Therefore, the eigenstate around the two valleys for the second and third bands can be well approximated by a linear combination of the degenerated eigenstates Eqns. (7) and (8). Thus an effectively continuum Hamiltonian in the neighbor of **K′** and **K** points can be derived by spanning the Eq. (3) over the degenerated states {$\mathbf{u}_1$(**K′**), $\mathbf{u}_2$(**K′**)} and {$\mathbf{u}_1$(**K**), $\mathbf{u}_2$(**K**)}, respectively. Following the procedure in **APPENDIX I**, the effective model can be expressed as

$$\Delta H\psi = \Delta\omega\psi \,, \tag{9}$$

$$\Delta H = m\sigma_z + v(\tau\Delta k_y\sigma_x - \Delta k_x\sigma_y)\,, \tag{10}$$

$$m = \frac{1}{2}\frac{m_q - m_p}{m_q + m_p}\omega_0, \quad v = \frac{\omega_0 a}{4\sqrt{3}}, \quad \omega_0 = \sqrt{\frac{3t}{m_p + m_q}}\,, \tag{11}$$

In the above equations, parameter $\tau = +1$ ($-1$) is introduced to denote the **K′**(**K**) valley, $\sigma_x$, $\sigma_y$ and $\sigma_z$ are the Pauli matrices, $\Delta H$ is the effective Hamiltonian and $\psi$ is an eigenvector composed of coefficients for the linear combination. Parameter $m$ and $v$ represents the Dirac mass and Dirac velocity, respectively. The relation **K′**(**K**) + $\Delta$**k** ~ $\omega_0$ + $\Delta\omega$ should be understood as the dispersion relation. From the effective model, eigen-frequencies and eigenstates can be obtained easily as

$$\Delta\omega = \pm\sqrt{m^2 + v^2\Delta k^2}, \quad \psi = \frac{\left(v(i\Delta k_x + \tau\Delta k_y),\ \Delta\omega - m\right)^{\mathrm{T}}}{\sqrt{(\Delta\omega - m)^2 + v^2\Delta k^2}}\,, \tag{12}$$

where the eigenstates are normalized according to $<\psi \mid \psi> = \psi^{\dagger}\ \psi = 1$ with a dagger in superscript $\psi$ represents conjugate transpose. It is seen that the higher and lower band are symmetric with respect to $\Delta\omega = 0$. Dirac degeneracy with a slope $v$ is obtained if the two masses are equal, otherwise a bandgap with width $|2m|$ occurs.

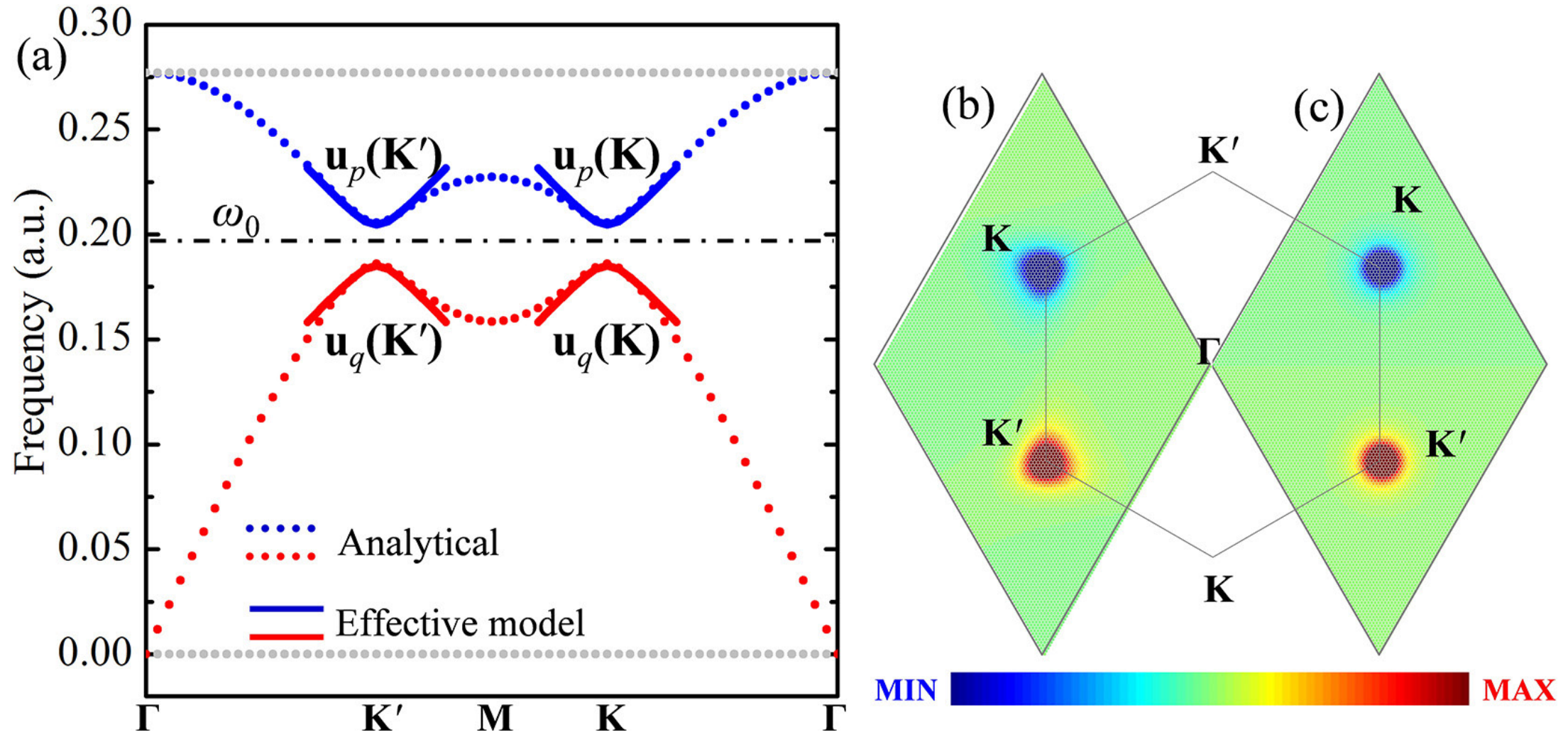

Figure 2: (a) Band structure for an unequal mass case ($m_p = 0.9$, $m_q = 1.1$, $t = 1$). (b) (c) Berry curvatures calculated from the discrete model and effective model, respectively.

Figure 2(a) shows band structure calculated from Eq. (6) for the case of an unequal mass ($m_p = 0.9$, $m_q = 1.1$, $t = 1$). Valley states predicted by Eqns. (7) and (8) are designated in the figure for clarity. The dispersion curves calculated from the perturbation model are also accompanied with solid curves. It is found that excellent approximation between the two sets of curves is achieved at the neighbor of valley, and this validates the effective model.

## 3. Valley Hall insulator and Chern insulator

The valley Chern number and Chern number can be used to characterize the valley Hall insulator and Chern insulator. They are obtained by integrating Berry curvature over a portion of the Brillouin zone or entire Brillouin zone. Berry curvature is defined from eigenstates, which can be obtained from the original eigenvalue equation Eq. (3) or from the $k \cdot p$ perturbative model Eq. (9). Here, we adopt eigenstates from the effective model Eq. (12) for mathematical simplification. Berry curvature for the lower branch around the valley can be derived,

$$F(\tau\mathbf{K}') = i\nabla_{\mathbf{k}} \times \langle \phi | \nabla_{\mathbf{k}} \phi \rangle = i\left(\frac{\partial \phi^{\dagger}}{\partial \Delta k_x}\frac{\partial \phi}{\partial \Delta k_y} - \frac{\partial \phi^{\dagger}}{\partial \Delta k_y}\frac{\partial \phi}{\partial \Delta k_x}\right) = \frac{\tau m v^2}{2(m^2 + v^2 \Delta k^2)^{3/2}} . \quad (13)$$

The calculated Berry curvatures together with the one from the discrete model Eq. (3) are compared by Figs. 2(b) and (c). The good agreement confirms again the effective model. Because $m$ is a small parameter, nonzero Berry curvature localizes around the two valleys and decays quickly to zero away from them. The valleys $\mathbf{K}'$ and $\mathbf{K}$ carry opposite Berry curvatures, hence the summation of Berry curvature over the two valleys gives a zero Chern number. This means the system is not a Chern insulator. However, the valley Chern number defined as integration over a single valley (Zhang, et al., 2013) could be non-trivial,

$$\mathrm{C}(\tau\mathbf{K}') = \frac{1}{2\pi}\iint F(\tau\mathbf{K}') d(\Delta k_x) d(\Delta k_y) = \frac{1}{2}\mathrm{sgn}(m)\tau . \quad (14)$$

i.e. the valley Chern numbers for the $\mathbf{K}'$ and $\mathbf{K}$ are +1/2 and −1/2, respectively. Thus the lattice becomes a valley Hall insulator with non-zero valley Chern number because of the reduction from $C_{6v}$ to $C_{3v}$ symmetry by the unequal masses $m_p \neq m_q$.

To realize Chern insulator in the above system, one needs to break the TR symmetry. This can be done by placing the entire system on a rotating platform or coupling a spinning top to each mass (Kariyado and Hatsugai, 2015; Wang, et al., 2015a). We here suppose the system is placed on a rotating platform with an angular frequency $\Omega$. In a coordinate system attached on the rotating platform, each mass experiences virtual centrifugal force and Coriolis force. For a small rotation speed $\Omega$, the Coriolis' force $\mathbf{F}_c = -(2\Omega\mathbf{e}_z) \times (m\mathbf{v})$ dominates and can be characterized by assigning each mass an effective mass matrix $m$ ($\mathbf{I}$ + $\gamma\sigma_y$) with $\gamma$ = $2\Omega/\omega$, $\mathbf{e}_z$ means out of plane unit vector. The virtual Coriolis' force depends on motion direction of the mass on the site and therefore breaks the TR symmetry. The effective Hamiltonian taking account of the Coriolis' effect is in the same form as Eqns. (9) - (11), except that the effective Dirac mass are updated as

$$m = m' + m''\tau, \quad m' = \frac{1}{2}\frac{m_q - m_p}{m_p + m_q}\omega_0, \quad m'' = \Omega\,. \tag{15}$$

If only the TR symmetry is broken, i.e., $m' = 0$ and $m'' \neq 0$, the effective masses $m$ for the two valleys are of opposite sign and a complete bandgap is formed between the second and third branches. Focusing on the lower branch and referring to Eqns. (13) and (14), because the Berry curvatures for the two valleys now are the same, the system becomes a Chern insulator with integer Chern number C = sgn($m''$). The rotating platform introduces a preferred intrinsic direction in the system, and interesting asymmetric wave propagation could be expected.

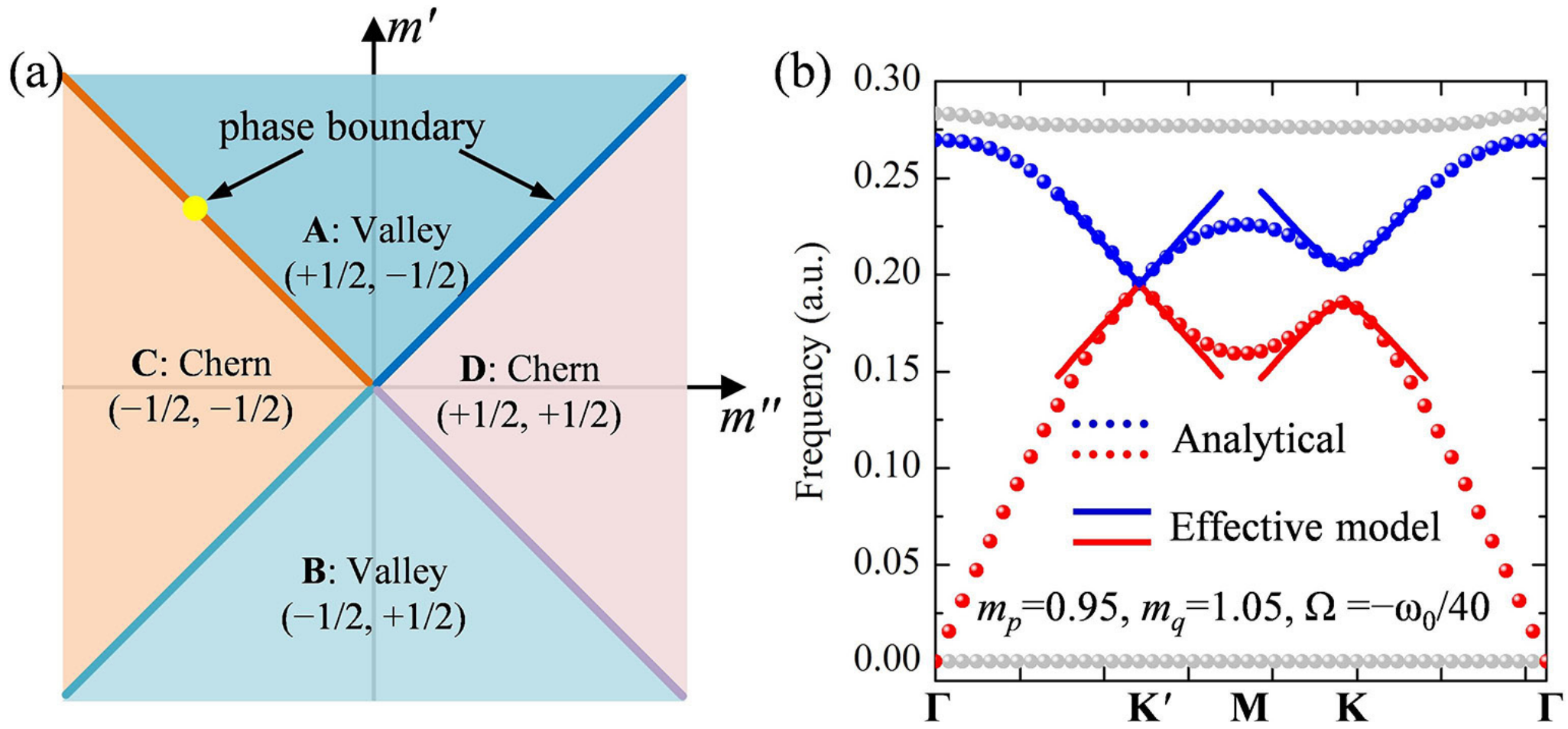

Figure 3: (a) Topological phase diagram; (b) Band structure for a case with parameters locating at intersection line of phase A and C, as indicated by the yellow point in (a).

When both the TR symmetry and inversion symmetry are broken, the system can be one of four phases according to the competition of breaking strength, as shown by the phase diagram in Fig. 3(a). Among the four phases, two of them are valley Hall insulators and the other two are Chern insulators. Numbers in brackets indicate the two valley Chern numbers for **K′** and **K**, and the Chern number is their summation. In the regions **A** – **D** shown in the parameter space (Fig. 3(a)), the system is either a valley Hall insulator or Chern insulator in the bulk with a complete bandgap. However, when the parameters of the system locate at intersection line of the four regions, the system is not an insulator with complete bandgap. The second and third branches are gapped only at one of the two valleys **K′** or **K**, while a Dirac cone is formed at the other valley as shown in Fig. 3(b) with system parameters $m_p$ = 0.95, $m_q$ = 1.05 and $\Omega = -\omega_0/40$. This characteristic occurs because the rotating platform has different effect on the gaps at the two valleys; the gap width at **K** is increased by the rotation while the gap width at **K′** is decreased. This feature clearly indicates a non-reciprocal wave phenomenon by breaking the TR symmetry.

### *3.1. Interface states between valley Hall and Chern insulators*

At domain walls separating different phases, topologically protected gapless interface states are guaranteed by the bulk-boundary correspondence (Bansil, et al., 2016; Hasan and Kane, 2010). With the $k \cdot p$ effective Hamiltonian, the existence, directionality, dispersion and mode of the interface wave can be analyzed conveniently. For illustration, here we consider a general interface along $y'$ direction formed by the A / B valley Hall insulators (Fig. 4, $x' < 0$, $m'' = 0$, $m' > 0$; $x' > 0$, $m'' = 0$, $m' < 0$), and the interface geometry is supposed to be periodic along the $y'$ direction (Fig.4(a)). The interface is of zig-zag shape for $\theta = 0^{\circ}$ (Fig. 4(b)), while it is of armchair shape for $\theta = 30^{\circ}$ (Fig. 4(c)). From the effective model Eq. (9), elastic wave in the lattices located at both sides of the interface is governed by the following differential equation (by mapping $\Delta k_x \rightarrow -i\partial_x$, $\Delta k_y \rightarrow -i\partial_y$),

$$m\sigma_z\psi - iv\tau\sigma_x\frac{\partial\psi}{\partial y} + iv\sigma_y\frac{\partial\psi}{\partial x} = \Delta\omega\psi \tag{16}$$

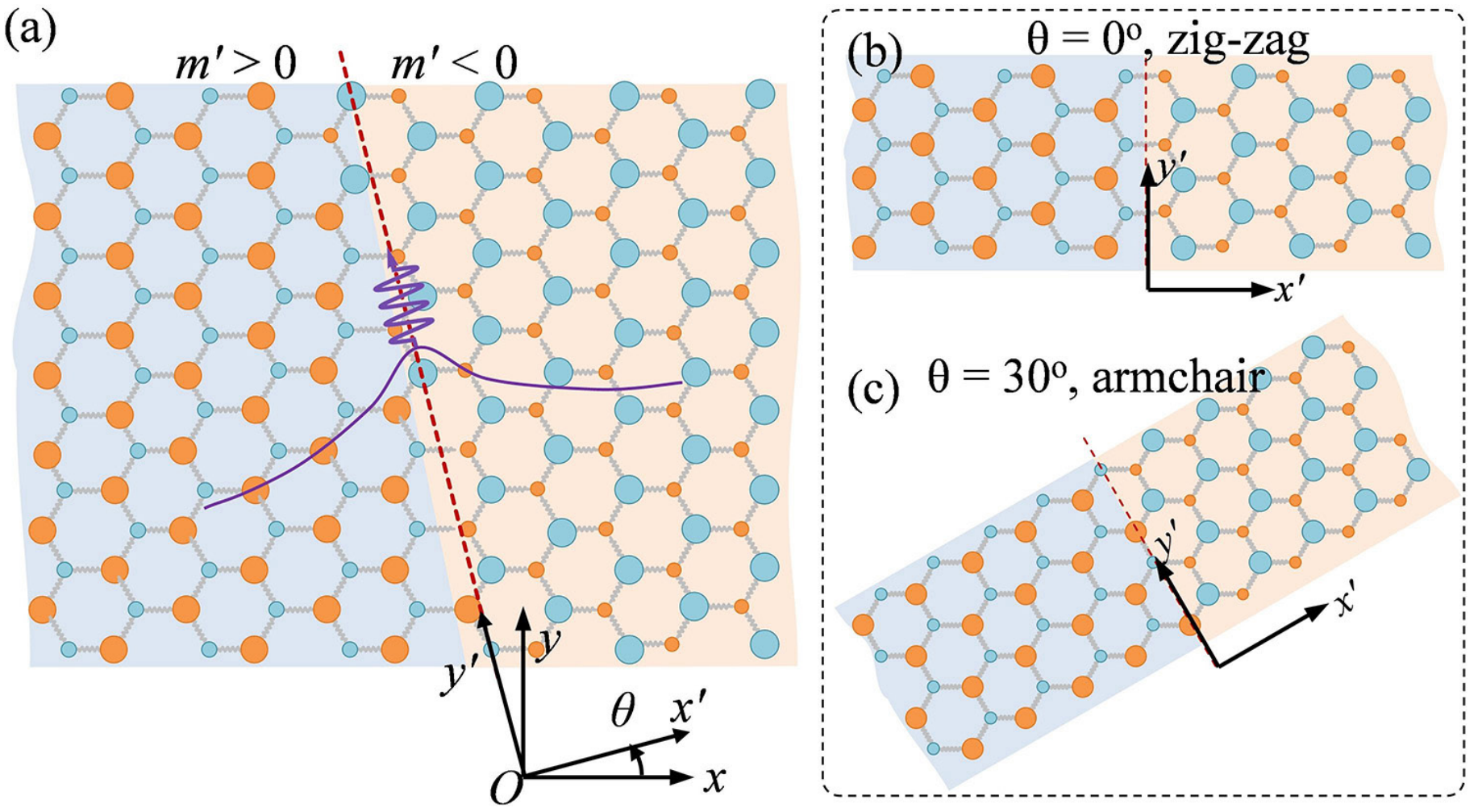

Figure 4: (a) Sketch of interface formed by distinct phases, interface shape depends on orientation angle $\theta$; (b)(c) Zig-zag / armchair interface corresponds to $\theta = 0^{\circ}$ / $30^{\circ}$.

To derive an interface solution for the periodic interface, a Bloch wave solution $\psi(x, y) = (c_1, c_2)^T \exp(\lambda' x') \exp(i\Delta k'_y y')$ is assumed, where $c_1$ and $c_2$ are unknown coefficients to be determined. Substituting the Bloch wave solution into Eq. (16) yields,

$$\begin{pmatrix} m & \exp(-i\tau\theta)v(\tau\Delta k'_y + \lambda') \\ \exp(i\tau\theta)v(\tau\Delta k'_y - \lambda') & -m \end{pmatrix}\begin{pmatrix} c_1 \\ c_2 \end{pmatrix} = \Delta\omega \begin{pmatrix} c_1 \\ c_2 \end{pmatrix} \tag{17}$$

Then eigen-frequency and eigenstate can be derived,

$$(\Delta\omega)^2 = m^2 + v^2\left((\Delta k'_y)^2 - \lambda'^2\right), \quad \frac{c_1}{c_2} = \exp(-i\tau\theta)\frac{v(\tau\Delta k'_y + \lambda')}{\Delta\omega - m} \tag{18}$$

To find interface solution localized to the interface, the parameter $\lambda' x'$ should be negative for both sides such that the solution decays exponentially away from the interface. Furthermore, the ratio $c_1/c_2$ must be continuous across the interface to ensure continuity condition. Since $\Delta\omega/\omega_0$ and $\Delta k'_y$ are the same at both sides, one can easily verify that the following exponentially decaying solution with linear dispersion always exists,

$$\psi = \frac{\sqrt{2}}{2}\begin{pmatrix} 1 \\ -\exp(i\tau\theta) \end{pmatrix}\exp(-|\frac{m'(x')}{v}x'|)\exp(i\Delta k'_y y'), \quad \Delta\omega = -\tau v \Delta k'_y \tag{19}$$

Substituting the valley states at $\mathbf{K}'$ and $\mathbf{K}$ into the above expression, displacement fields for the sites $p$ and $q$ is derived,

$$\begin{aligned} \mathbf{u} &= \left(\mathbf{u}_1(\tau\mathbf{K}')\exp(i\tau\mathbf{K}'\cdot\mathbf{r}),\ \ \mathbf{u}_2(\tau\mathbf{K}')\exp(i\tau\mathbf{K}'\cdot\mathbf{r})\right)\cdot\psi(x,y) \\ &= \frac{1}{2}(-i\tau, 1, -i\tau\exp(i\tau\theta), -\exp(i\tau\theta))\exp(i\tau\mathbf{K}'\cdot\mathbf{r})\exp(-|\frac{m'(x')}{v}x'|)\exp(i\Delta k'_y y') \end{aligned} \tag{20}$$

It is seen that the interface states are simply linear combinations of the previous valley states supplemented by a decaying factor and a phase term. From the dispersion relation Eq. $(19)_2$, the solution formed by the valley states of $\mathbf{K}'$ propagates along $-y'$ direction, while the one formed by that of $\mathbf{K}$ propagates along the opposite direction. This correspondence will reverse if the adjacent valley Hall insulators exchange their position. The decaying factor $|m(x')/v|$ depends on the effective mass, the larger the mass difference is, the more the interface state is localized. In addition, the interface state exists for arbitrary angle $\theta$, which implies that elastic wave can propagate along arbitrarily oriented interface, and this proves the topological robustness of the interface state. However, the above theoretical model does not include directly the influence of changes in angle along the path. In following studies, we will numerically demonstrate the interface state is also robust to changes in angle.

Figures 5(a) and 5(b) show a strip with an interface by the A / B phases (A, $m_p = 0.9$, $m_q = 1.1$, $\Omega = 0$; B, $m_p = 1.1$, $m_q = 0.9$, $\Omega = 0$) and the corresponding numerically solved band structure, respectively. The interface is along the $y$ direction with $\theta = 0^{\circ}$ according to the definition in Fig. 4(b). A gapless blue curve in the bulk gap is observed and represents the above derived interface solution. This can also be verified from the plotted displacement fields (Figs. 5(c) − (e)) for the three indicated states for the valley $\mathbf{K}'$. For the interface state, only sites close to the interface have a strong vibration (Fig. 5(d)). The interface mode at the valley $\mathbf{K}'$ has a negative group velocity as theoretically predicted in Eq. (19).

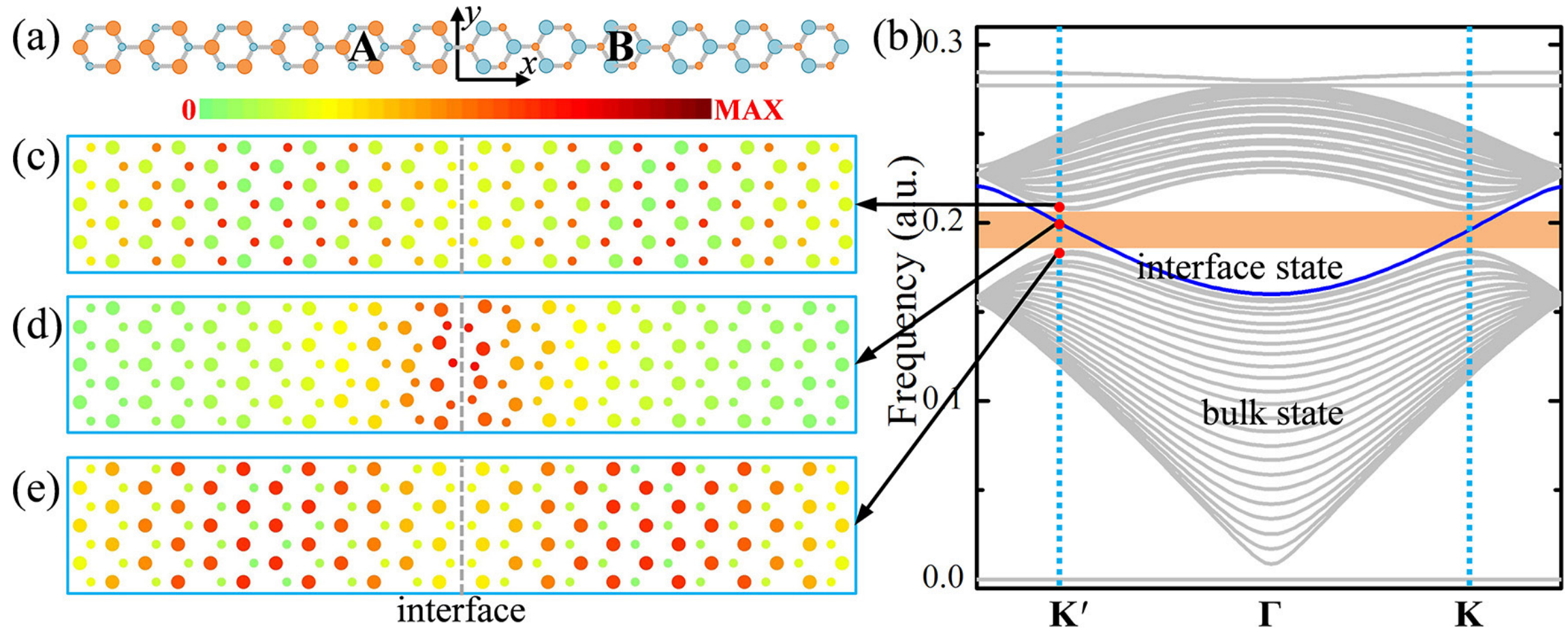

Figure 5: (a) Strip geometry with an interface formed by A / B phases; (b) Band structure of the strip geometry obtained by setting a Bloch wave condition along the $y$ direction; (c) (e) Displacement magnitudes for the indicated upper and lower bulk states; (d) Displacement magnitude for the indicated interface state; Red color means strong vibration.

Figure 6(b) shows vibration trajectory for sites close to the interface. For the interface state at $\mathbf{K}'$, the site $p$ / $q$ rotates counter-clockwise / clockwise around the equilibrium position, in consistent with the above analytical solution. For the other interface state at the valley $\mathbf{K}$, all sites reverse their rotation direction. Since the $k \cdot p$ effective continuum model is isotropic, the analytical solution Eq. (20) predicts a circular trajectory for all the sites. However, the actual trajectory is an ellipse with principal axis along the $x$ or $y$ direction. This result is obtained because the honeycomb lattice has a $C_{3v}$ symmetry instead of being completely isotropic. Distribution of horizontal displacement for sites $q$ / $p$ along the $x$ direction is shown in Figs. 6(d)(e), respectively. The displacement decays exponentially away from the interface and agrees well with the prediction of Eq. (20).

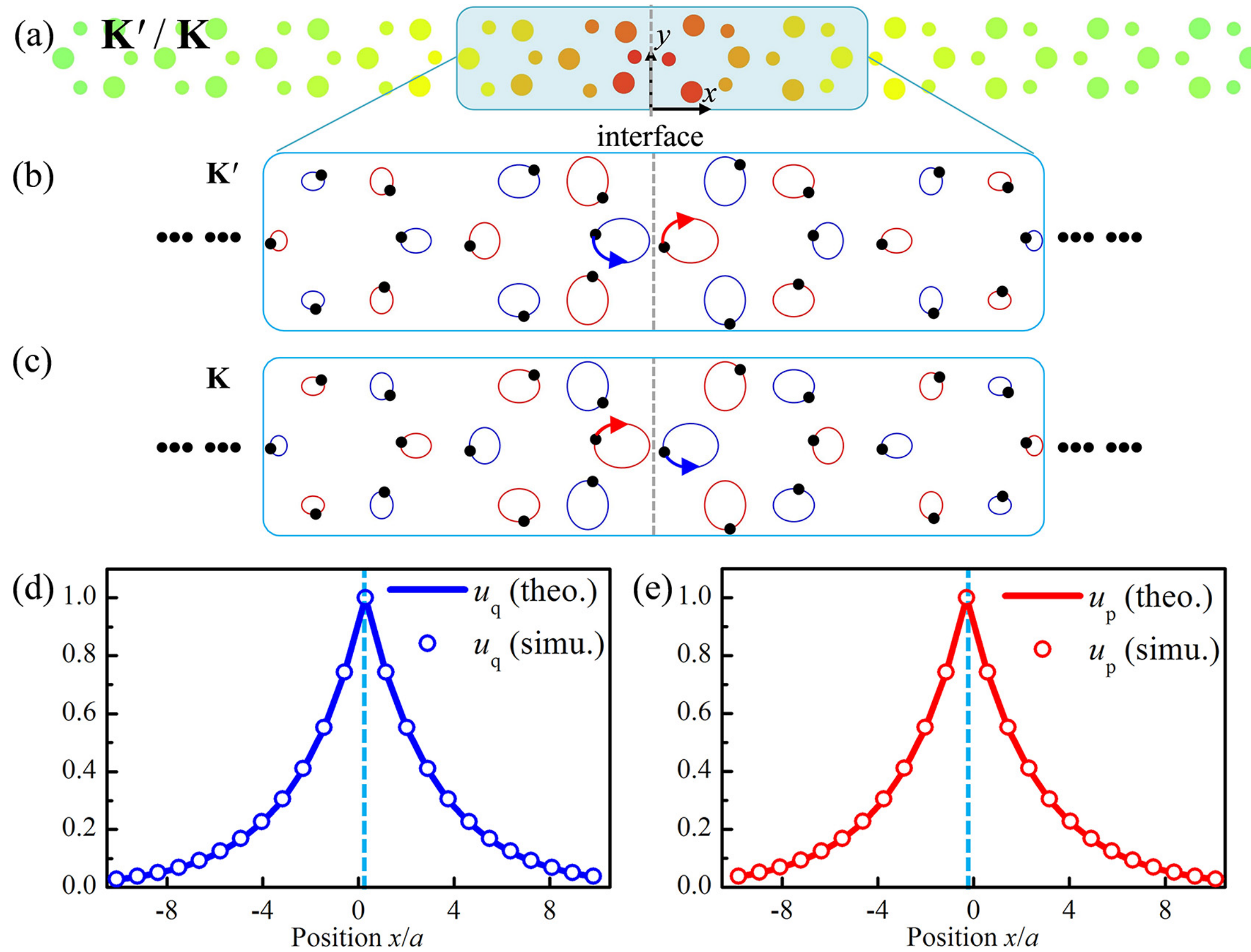

Figure 6: (a) Displacement magnitude of sites for interface state at **K**(**K′**). (b) (c) Trajectory of sites close to the interface, black dot indicates the site position; Red / Blue means clockwise / counter-clockwise rotation along the trajectory; (d) (e) Horizontal displacements for sites $q$ / $p$ along the $x$ direction.

Interface states for combinations of the other phases can be derived similarly. For instance, at interface between two Chern insulators with Chern number −1 and +1 ($x' < 0$, $m' = 0$, $m'' < 0$; $x' > 0$, $m' = 0$, $m'' > 0$), the interface solution is,

$$\psi(x,y)=\frac{\sqrt{2}}{2}\begin{pmatrix}1\\ \tau\exp(i\tau\theta)\end{pmatrix}\exp(-|\frac{m''(x')}{v}x'|)\exp(+i\Delta k'_y y'),\qquad \Delta\omega=-v\Delta k'_y \tag{21}$$

Since the Chern number is differed by a value two across the interface, two interface solutions are there, and are also composed of the above valley sates at **K′** and **K**. However, here the two states propagate both along the $+y'$ direction. The propagating direction will be reversed if the two Chern insulators exchange their positions.

Figures 7 (a) – (f) shows band structures for 6 types of zig-zag interface through combination of the above four phases (A, $m_p = 0.9$, $m_q = 1.1$, $\Omega = 0$; B, $m_p = 1.1$, $m_q = 0.9$, $\Omega = 0$; C, $m_p = 1.0 = m_q$, $\Omega = -\,0.05\omega_0$; D, $m_p = 1.0 = m_q$, $\Omega = +\,0.05\omega_0$). The band structures are solved numerically based on the same geometry as in Fig. 4(b). Interface band marked by blue line is observed in the bulk bandgap. The interface between two Chern insulators C / D supports two interface states, while interface between the other phases only carries a single interface state. The interface between A / B valley Hall phases supports bi-directional propagating state, while the interface state for Chern insulators only travel along one direction (Figs. 7(b) – (f)) as previous analysis. It should be noted here that, because the free space can be considered as trivial insulator, interface state also emerges at an interface between the free space and a Chern insulator. Such an interface state is often referred as edge state since the free space could not support vibration. The edge states localized on edges of Chern insulators are indeed observed in the simulation, however are not plotted here.

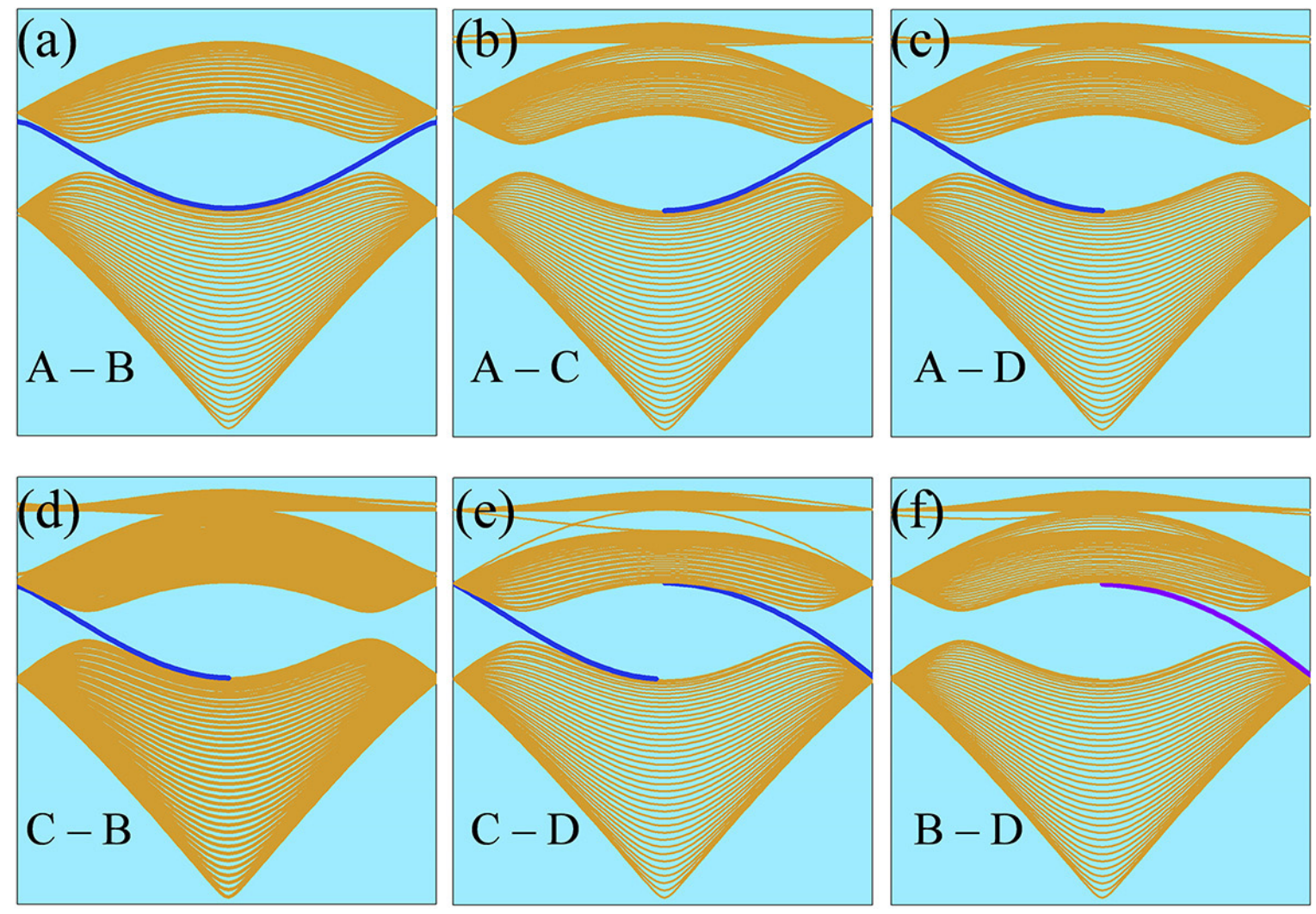

Figure 7: (a) – (f) Band structure for strip geometry with a zig-zag interface formed by different phase combinations; Brown / Blue color is for the bulk / interface band.

Transient wave propagation along a designed interface is also simulated to validate the above interface states (Fig. 8), where red color stands for large displacement amplitude. The interface is formed along the boundary of two regions occupied by two different phases, a central diamond shaped region and an exterior region. Some more sharp bending angles, e.g., 110° or 80° are intentionally designed along the interface. Such a complex interface can be used to demonstrate more efficiently topological robustness of the interface states. In the transient simulation, displacement for the $p$ and $q$ sites in a unit cell at the indicated location is set according to the valley states at $\mathbf{K}'$, $(\mathbf{u}_1(\mathbf{K}') + \mathbf{u}_2(\mathbf{K}')) \times \exp(-(\omega_0 t/120)^2) \times \cos(\omega_0 t)$, where the excitation frequency $\omega_0 = 2\pi/T_0$ lies in the center of bulk gap. The resulted Gaussian pulse has a time duration about $75T_0$. It is observed that, the elastic wave can flow in both directions (Fig. 8(a)) along the interface of A / B valley Hall phases, while only travels in one direction along the interfaces with Chern insulators (Figs. 8(b) – (f)) (see Supplemental materials for animation). All these agree with the previous analytical solution and band structure, and the stimulated wave is able to transmit perfectly along the interface without evident backward reflection or scattering induced by sharp corners. We remark here, because the interface modes, projected from the **K**′ and **K** valley, of valley Hall insulator propagate along opposite directions, it is less robust than the unidirectional interface mode(s) of Chern insulator. Incident wave along the interface of valley Hall insulator in principle may be reflected by the inter-valley scattering (Lu et al., 2013), while the backward scattering is strictly forbidden for the interface of Chern insulator. However, the reflection caused by inter-valley scattering is usually at the level of 1% (Lu et al., 2016). Therefore, the wave transmission along the valley Hall interface (Fig. 8(a)) is as efficient as that along the Chern interface (Figs. 8(b)–(f)). The simulation demonstrates a very high transmission efficiency and robustness of the interface modes, which are almost impossible in conventional waveguides.

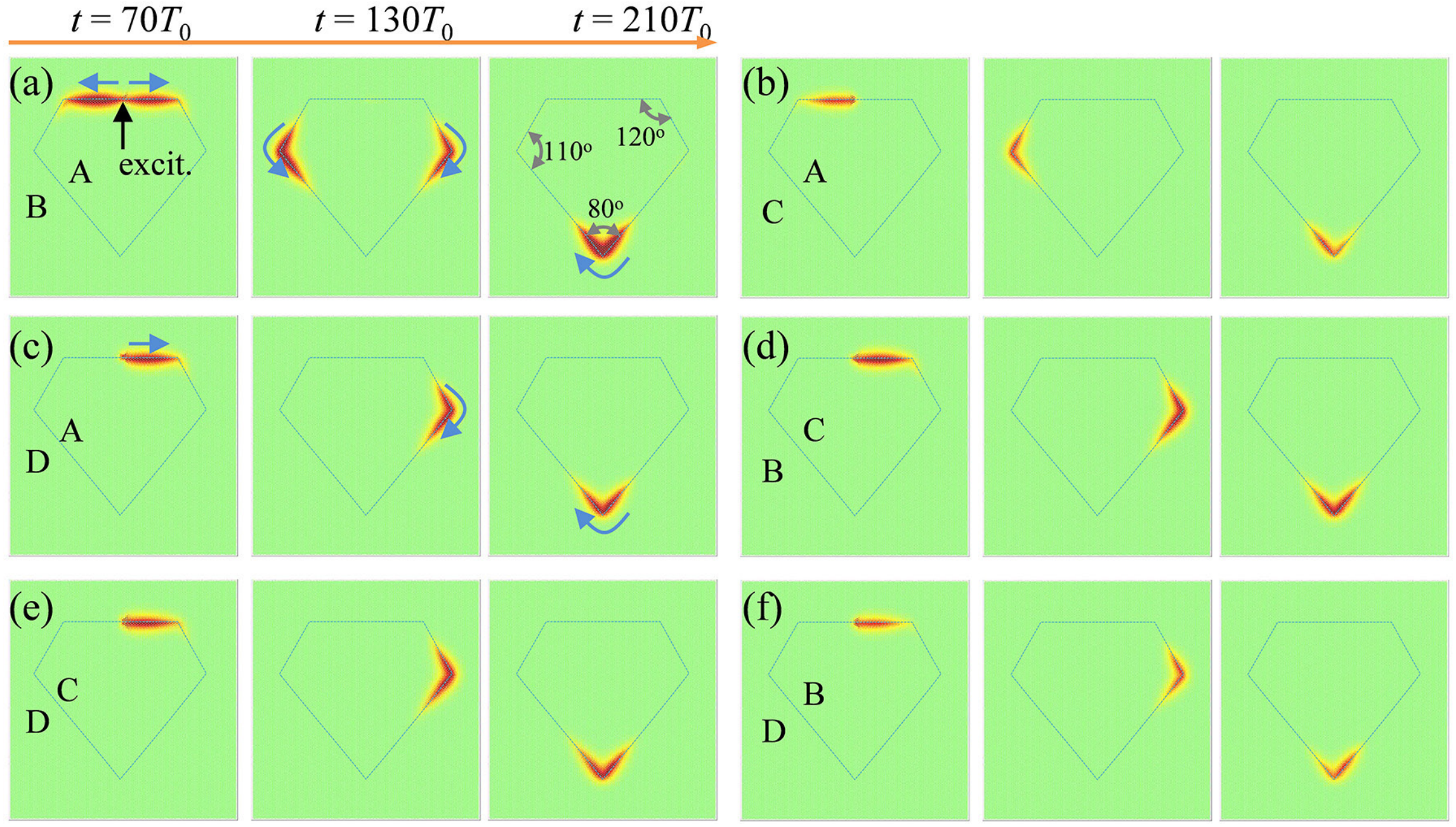

Figure 8: (a) – (f) Snapshot of simulated elastic wave propagating along the above 6 types of interface; Color is for displacement magnitude of sites in the lattice; Red means large displacement amplitude. The simulation domain is composed of 78 × 90 unit cells.

## 4. Spin Hall insulator

In the previous sections, passive QVHE insulator is achieved by tuning the mass to break inversion symmetry, while the Chern insulator always requires active means to break TR symmetry. In this section, we present in the same lattice a way to realize another passive non-trivial phase, the QSHE insulator, by tuning the spring constants. For the genuine QSHE insulator, the spin-orbit coupling of electrons is invoked for non-trivial topology, and protected helical edge states with opposite chirality and transport directions are supported for spin-up and spin-down cases, respectively (Kane and Mele, 2005a). To reproduce the effective spin-orbit coupling in a classical wave system, a key point is to construct the fictional pseudo-spin state. To this end, we will adopt the supercell method proposed initially for photonic system by Wu and Hu (2015) to design the QSHE insulator within the same mechanical honeycomb lattice examined in the previous sections.

As shown in Fig. 9(a), in the honeycomb lattice with uniform mass $m_0 = 1$, a super cell is created by grouping a mass cluster with alternating spring stiffness. The unit cell is indicated by the hexagon, in which the masses are numbered from 1 to 6. Sites in the same cell are linked by intra-cell springs with stiffness $t_i$, while sites in different cells are linked by inter-cell springs with stiffness $t_o$. The first BZ of the lattice is shown in Fig. 9(b) as the small grey hexagon. For the special case $t_i = t_o$, the rhombus in Fig. 9(a) should be regarded as the unit cell, and the large green hexagon in Fig. 9(b) is the corresponding first BZ. In this case, Dirac cones will emerge at the BZ corners $\mathbf{K}'$ and $\mathbf{K}''$. Since $\mathbf{\Gamma K}'$ and $\mathbf{\Gamma K}''$ are reciprocal vectors of the supercell, the two Dirac cones will be folded into Brillouin zone center $\mathbf{\Gamma}$ to form a double fold Dirac cone. The pseudo-spin state and non-trivial QSHE phase can be realized by tuning the spring stiffness to lift the double fold Dirac cones.

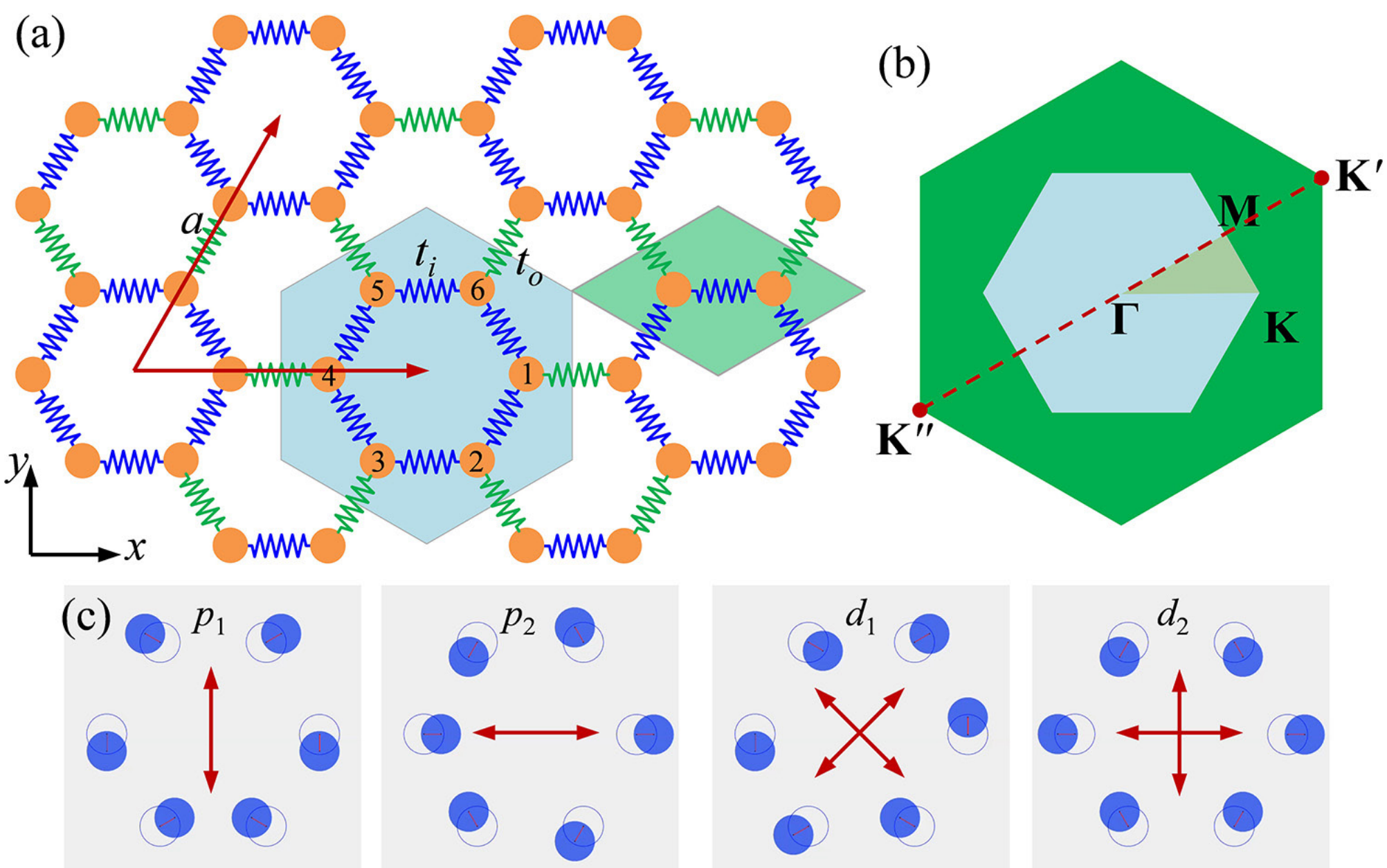

Figure 9: Double fold Dirac cone by band folding; (a) Supercell composed of 6 sites; (b) Small hexagon is Brillouin zone for the supercell; Green hexagon is Brillouin zone of a honeycomb lattice with uniform springs. (c) The $p$ and $d$ states at $\mathbf{\Gamma}$; Solid blue circle denotes location of maximal vibration; hollow circle indicates equilibrium position.

Each unit cell with six sites has in total 12 displacement DOFs. An eigenvalue problem of dimension 12 can therefore be derived for Bloch wave in the lattice as before. The mass and stiffness matrices are not listed here due to quite lengthy expression. The eigenvalue problem results in total 12 frequency bands, while we only focus on the degenerated bands at **Γ**. For general case $t_i \neq t_o$, the 5$^{th}$ to 8$^{th}$ bands form a pair of doubly degenerated bands at **Γ**, while a double fold Dirac cone with the following eigenstates can be derived if we have equal stiffness $t_i = t_o = t$,

$$\omega^2 = \frac{3t}{2m_0}, \quad \begin{cases} |p_1\rangle = \{0,-2,-\sqrt{3},1,\sqrt{3},1,0,-2,-\sqrt{3},1,\sqrt{3},1\}^{\mathrm{T}} \\ |p_2\rangle = \{2,0,-1,-\sqrt{3},-1,\sqrt{3},2,0,-1,-\sqrt{3},-1,\sqrt{3}\}^{\mathrm{T}} \\ |d_1\rangle = \{0,2,-\sqrt{3},1,-\sqrt{3},-1,0,-2,\sqrt{3},-1,\sqrt{3},1\}^{\mathrm{T}} \\ |d_2\rangle = \{2,0,1,\sqrt{3},-1,\sqrt{3},-2,0,-1,-\sqrt{3},1,-\sqrt{3}\}^{\mathrm{T}} \end{cases} \tag{22}$$

The above four eigenstates $|p_1\rangle$, $|p_2\rangle$, $|d_1\rangle$, $|d_2\rangle$ give the displacements in $x$ and $y$ directions of the mass 1 to 6 in the unit cell. Because all the masses have a real valued displacement, the site vibrates back and forth through their equilibrium location (Fig. 9(c)). The $p$ states are anti-symmetric under spatial inversion operation, while the $d$ states are symmetric. This symmetry pattern is exactly the same as the $p$ / $d$ atomic orbital wave function for electrons, and the eigenstates are therefore named accordingly. The symmetry and degeneracy property of the $p$ / $d$ states originates from the point and space symmetry group of the lattice, i.e. the $C_{6v}$ symmetry group, which contains two two-dimensional irreducible representations. Despite the mechanical system has no intrinsic spin DOF, one can construct the pseudo spin states and the corresponding pseudo TR operator by observing the symmetry property of the $p$ / $d$ states (Wu and Hu, 2015),

$$T' = i\sigma_z K, \quad |p^{\pm}\rangle = (|p_1\rangle \pm i|p_2\rangle)/\sqrt{2}, \quad |d^{\pm}\rangle = (|d_1\rangle \pm i|d_2\rangle)/\sqrt{2} \tag{23}$$

Here, symbol + / – represents the pseudo spin-up / spin-down state, respectively, $K$ stands for the complex conjugate operator. For the pseudo spin-up / spin-down states combined with the $p$ states, the sites in a unit cell rotate counter-clockwise / clockwise around their equilibrium. While for the pseudo spin-up / down state constructed from the $d$ states, the sites rotate in the reversed direction. One can verify that, when the pseudo TR operator acts on the pseudo spin states, the spin direction will be reversed with an additional 180$^{\circ}$ phase, just like the TR operator acting on the electron spin. Using the pseudo spin states as expansion basis $\{|p^+\rangle, |d^+\rangle, |p^-\rangle, |d^-\rangle\}$, an effective model can be derived,

$$\Delta H \psi = \Delta\omega\psi \tag{24}$$

$$\Delta H = \begin{pmatrix} h^+(\mathbf{k}) & 0 \\ 0 & h^-(\mathbf{k}) \end{pmatrix}, \; h^{\pm}(\mathbf{k}) = (-\delta t + B\Delta k^2)\sigma_z - A\Delta k_x \sigma_y \pm A\Delta k_y \sigma_x \tag{25}$$

$$\delta t = \frac{t_o - t_i}{4m_0\omega_0}, \quad B = \frac{a^2 t_o}{16m_0\omega_0}, \quad A = \frac{at_o}{8m_0\omega_0}, \quad \omega_0 = \sqrt{\frac{2t_i + t_o}{2m_0}}, \tag{26}$$

where $h^+(\mathbf{k})$ / $h^-(\mathbf{k})$ represent the Hamiltonian for the pseudo spin-up / down state, $\psi$ is the eigenvector spanning over the pseudo spin basis. Parameters $\delta t$, $B$ and $A$ are the Dirac mass, spin-orbit coupling parameter and Dirac velocity, respectively. Here the spin-orbit coupling effect is controlled by the spring stiffness. The signs of $\delta t$ and $B$ in together dictate the topological phase transition in the system and can be quantitatively characterized by the spin Chern number. The pseudo spin-up and spin-down Hamiltonian are decoupled, and can be solved separately using $h^+(\mathbf{k})\,\psi^+ = \Delta\omega\,\psi^+$ and $h^-(\mathbf{k})\,\psi^- = \Delta\omega\,\psi^-$. We can then obtain a pair of doubly degenerated bands,

$$\Delta\omega = +\sqrt{(\delta t - B\Delta k^2)^2 + A^2\Delta k^2}, \psi^{\pm} = \frac{\left(iA(\Delta k_x \mp i\Delta k_y), \delta t - B\Delta k^2 + \Delta\omega\right)^{\mathrm{T}}}{\sqrt{A^2\Delta k^2 + (\delta t - B\Delta k^2 + \Delta\omega)^2}} \tag{27}$$

$$\Delta\omega = -\sqrt{(\delta t - B\Delta k^2)^2 + A^2\Delta k^2}, \psi^{\pm} = \frac{\left(iA(\Delta k_x \mp i\Delta k_y), \delta t - B\Delta k^2 + \Delta\omega\right)^{\mathrm{T}}}{\sqrt{A^2\Delta k^2 + (\delta t - B\Delta k^2 + \Delta\omega)^2}} \tag{28}$$

The lower and upper bands both consist of spin-up and spin-down band denoted as $\psi^+$ / $\psi^-$, the basis state of $\psi^+$ and $\psi^-$ are $\{|\,p^+\rangle,\ |\,d^+\rangle\}$ and $\{|\,p^-\rangle,\ |\,d^-\rangle\}$, respectively. The band gap with width $|2\delta t\,|$ at $\mathbf{\Gamma}$ depends on the spring stiffness and it closes when $t_i = t_o$. Here we will focus on the lower band as usual; the Berry curvature for the pseudo spin-up / down state is evaluated as

$$F^{\pm}(\mathbf{\Gamma}) = i\nabla_{\mathbf{k}} \times \left\langle \psi^{\pm} \,|\, \nabla_{\mathbf{k}}\psi^{\pm} \right\rangle = \mp \frac{A^2(\delta t + B\Delta k^2)}{2\left((\delta t - B\Delta k^2)^2 + A^2\Delta k^2\right)^{3/2}}. \tag{29}$$

The pseudo spin-up / down state carries opposite Berry curvature. The spin Chern number is defined as integration of the Berry curvature for each individual spin state,

$$\mathrm{C}^{\pm} = \frac{1}{2\pi}\iint_{\Omega} F^{\pm}(\mathbf{\Gamma})\, d\Delta k_x d\Delta k_y = \mp\frac{1}{2}(\mathrm{sgn}(\delta t) + \mathrm{sgn}(B)). \tag{30}$$

Summation of both spin Chern numbers for the up and down spin states is always zero, in consistent with the fact that TR symmetry is preserved. However, the spin Chern number, which characterizes the spin Hall phase transition, can be non-trivial. If the parameters $B$ and $\delta t$ have the same signs $B \times \delta t > 0$, the lattice is a spin Hall insulator with spin Chern number

±1; otherwise it is a trivial phase. Because the parameter $B$ is always positive $B > 0$, QSHE phase transition occurs when the system passes through the critical point of $t_i = t_o$, in which case the band gap closes. Spin Hall insulator is obtained if the inter-cell stiffness is larger than the intra-cell ones, $t_i < t_o$.

Band structure for three cases ($t_i = 1.05 > t_o = 0.90$; $t_i = 1.0 = t_o$; $t_i = 0.95 < t_o = 1.10$) are shown in Fig. 10, where color indicates the weight of the mixing of $p$ and $d$ states. The predicted double fold Dirac cone at $\mathbf{\Gamma}$ is observed for case of equal spring stiffness $t_i = t_o$, and split into upper and lower doubly degenerated bands for unequal cases. For the trivial case $t_i > t_o$ (Fig. 10(a)), $d$ state occupies the lower band, while for a QSHE insulator $d$ states occupy the higher band (Fig. 10(c)). Such a band inversion is also a hallmark of topological phase transition.

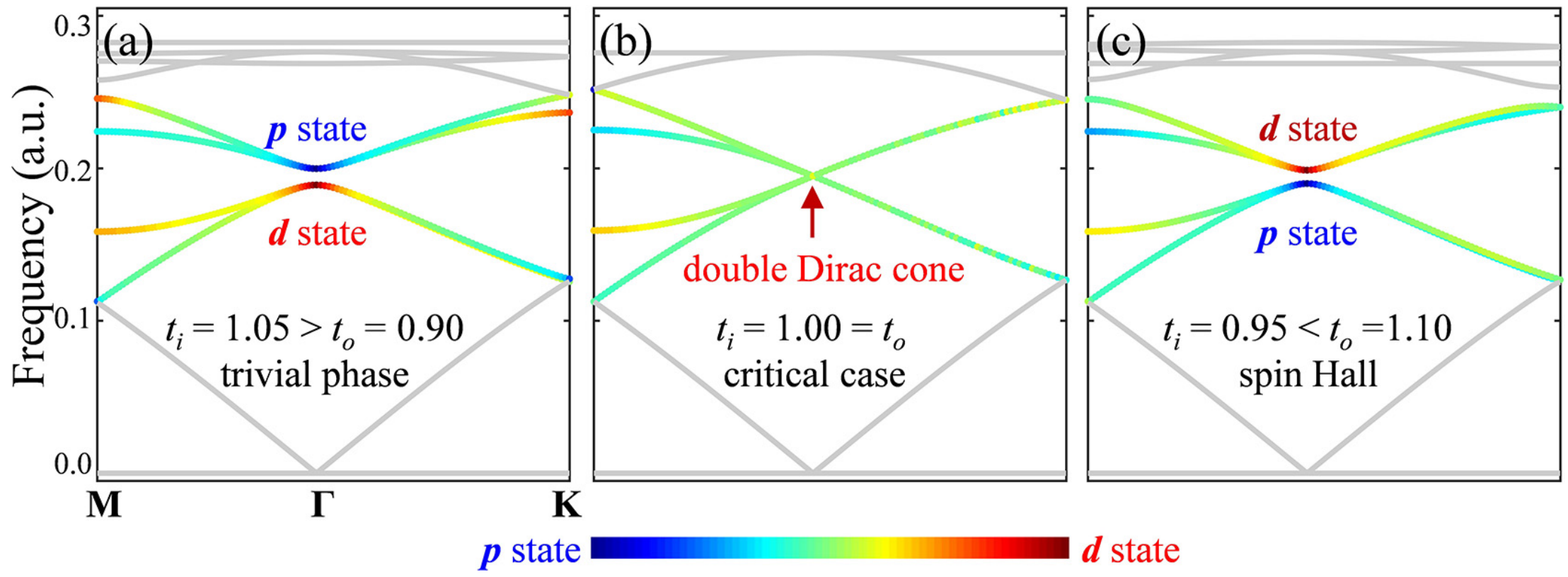

Figure 10: Quantum spin Hall phase transition; (a) (b) (c) Band structures for trivial insulator, critical case and spin Hall insulator, respectively; Rainbow color indicates hybridization of the states.

### 4.1. Spin-dependent interface states

To demonstrate spin controlled interface states, we consider an interface separating a trivial insulator and a spin Hall insulator ($x' < 0$, $\delta t < 0$; $x' > 0$, $\delta t > 0$) similar to the case in Fig. 4. To simplify the analysis, the effective model Eq. (25) is used by neglecting the second order spin-orbit coupling effect. This first order approximation is valid if a small wave vector $|\mathbf{k}| << 1$ is considered. Two degenerated interface states can be derived similarly,

$$\psi^{\pm}(x,y) = \frac{1}{\sqrt{2}}\begin{pmatrix} -1 \\ +\exp(\pm i\theta) \end{pmatrix} \exp(-|\frac{\delta t(x')}{A} x'|) \exp(i\Delta k'_y y'), \quad \Delta\omega = \mp A\Delta k'_y. \tag{31}$$

The spin-up interface state $\psi^+$ is composed of the pseudo spin-up state $|p^+\rangle$ and $|d^+\rangle$, while the spin-down mode $\psi^-$ is composed of the pseudo spin-down state $|p^-\rangle$ and $|d^-\rangle$. The two spin polarized interface states travel in opposite directions with the same group velocity $A$. Unidirectional upward (downward) propagating interface wave in the lattice is possible if one selectively excites the pseudo spin-down (spin-up) state. If the adjacent spin Hall insulator and trivial insulator exchange their position, this spin-direction correspondence will reverse.

Band structure for strip geometry (Fig. 11(a)) is numerically determined to demonstrate the counter-propagating interface modes. An interface along the $y$ direction is formed by a spin Hall insulator ($t_o$ = 1.10, $t_i$ = 0.95) in the left domain and a trivial phase ($t_o$ = 0.90, $t_i$ = 1.05) in the right domain, and the interface angle is $\theta = 0^o$. Two interface modes with opposite group velocity are clearly observed in the bulk gap (Fig. 11(b)), and large displacement in the lattice is localized at the interface (Fig. 11(c)). For the pseudo spin-up interface state, the displacement is almost linearly polarized as predicted by Eq. (31) and the sites rotate clockwise around their equilibrium positions. For the pseudo spin-down state, the sites rotate in reversed direction. The pseudo spin-up or down state can thus be selectively excited by rotating the sites accordingly.

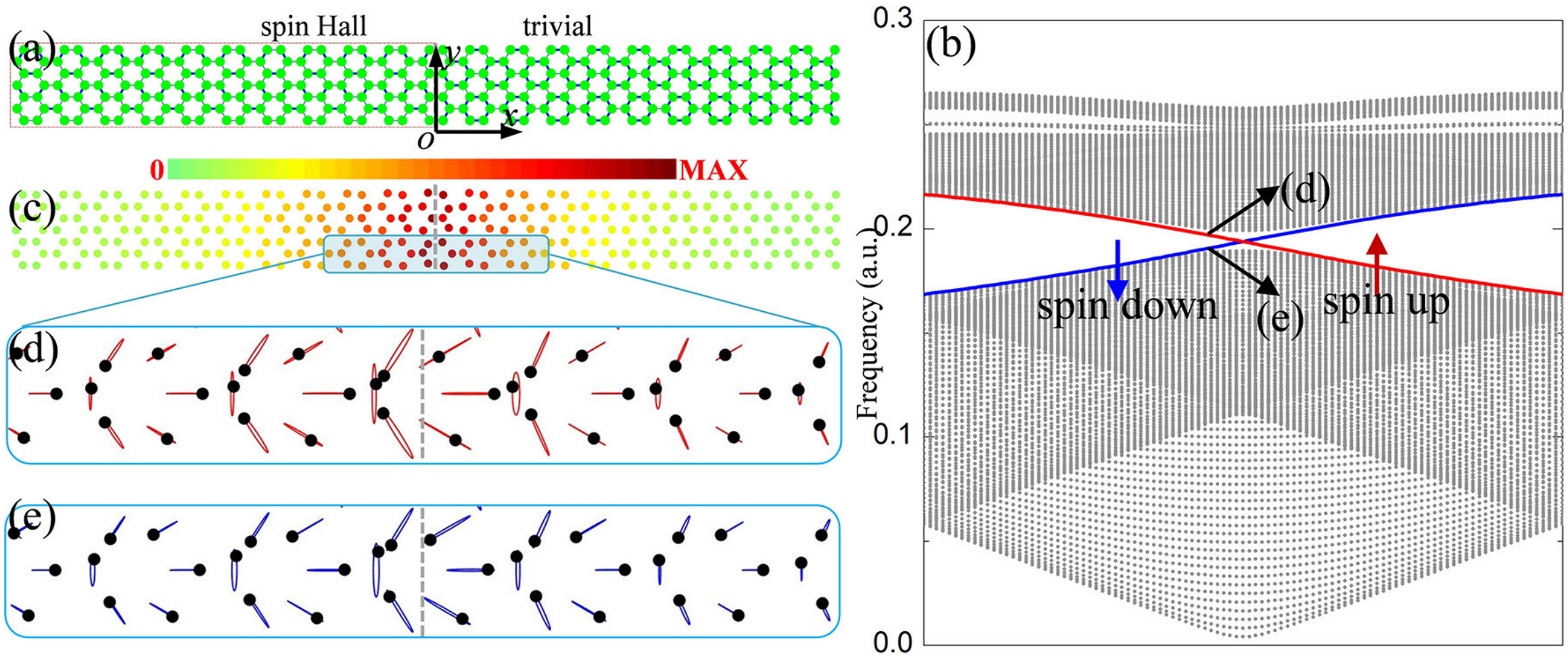

Figure 11: (a) Strip geometry with an interface between a spin Hall insulator and a trivial insulator. (b) Band structure of the strip; Gray is for bulk band; Red / blue indicates spin-up / spin-down interface band; (c) Displacement magnitude of the sites for the spin-up / down interface states. (d) (e) Trajectory of the sites close to the interface for the spin-up / down interface states; Red / blue indicates clockwise / counter-clockwise rotation.

It should be noted that the above gapless interface state occurs between an interface separating a QSHE insulator and a trivial insulator. Just like the electronic QSHE insulator, the mechanical QSHE insulator here also supports spin-polarized edge states on its edge. However, the edge states are not as robust as that in electronic system with intrinsic spin degree of freedom. The above pseudo spin states are constructed from Bloch states of the super cell with specific $C_{6v}$ symmetry group, any deviation from the $C_{6v}$ symmetry group can damage the pseudo spin edge states. Boundary of the lattice must be properly truncated to preserve a similar local spatial symmetry, otherwise the edge band may be pushed to the bulk band range and a complete bandgap will be observed in the band structure (Kariyado and Hu, 2017).

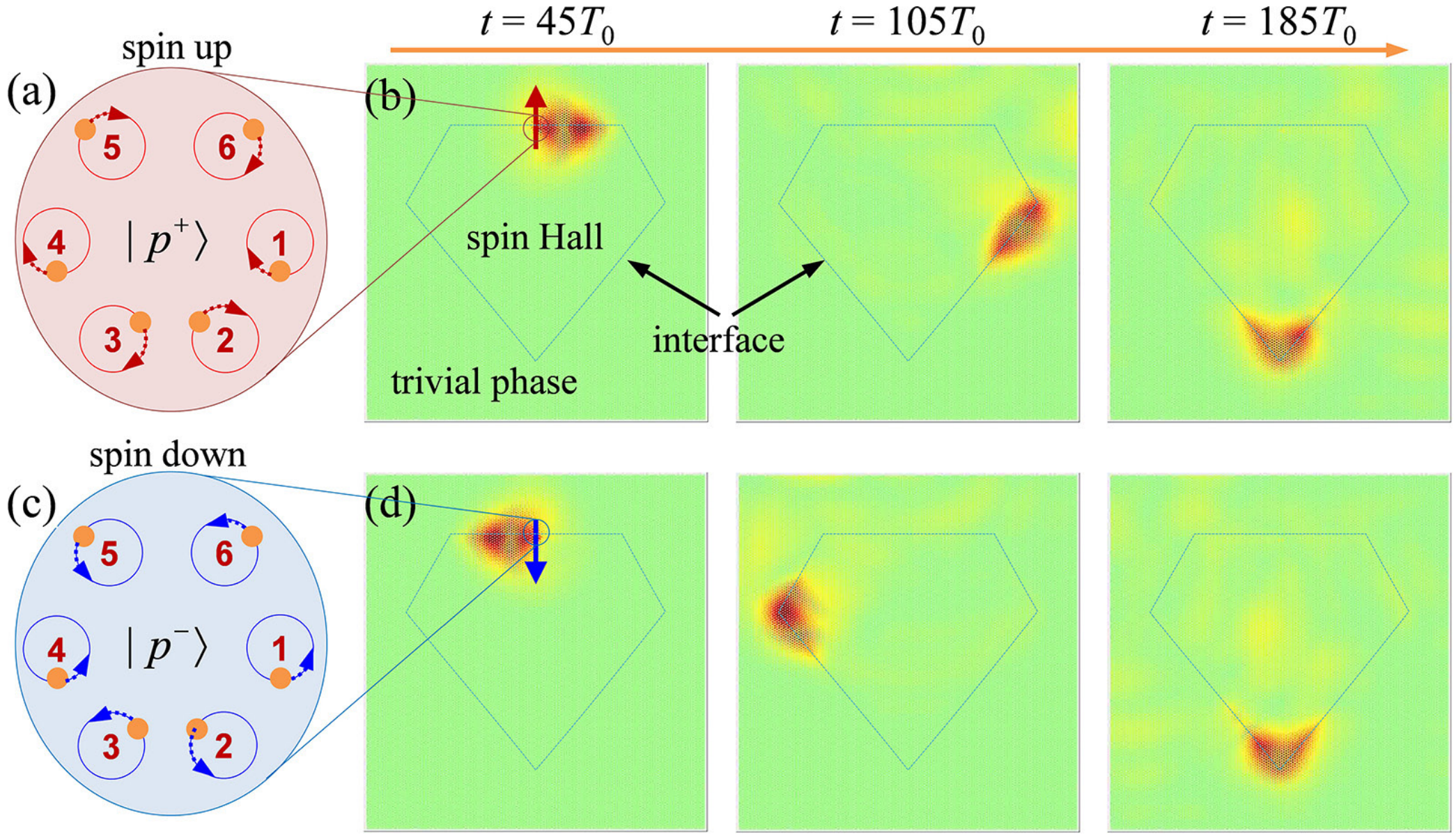

Figure 12: Unidirectional interface states of spin Hall insulator; (a) (c) Trajectory of the sites for the constructed pseudo spin-up / down excitation; (b) (d) Snapshots of simulated elastic wave propagation with pseudo spin-up / down state excitation; Red color indicates large displacement magnitude. The simulation domain is composed of 78 × 90 unit cells.

Figure 12 shows snapshots of simulated elastic wave propagation along the interface between a QSHE insulator ($t_o = 1.10$, $t_i = 0.95$, $m_0 = 1$) and a trivial lattice ($t_o = 0.90$, $t_i = 1.05$, $m_0 = 1$). In the simulation, transient displacements for six sites in a unit cell at indicated location are prescribed according to the pseudo spin state $|p^+\rangle$ or $|p^-\rangle$ with a Gaussian pulse modulation $\exp(-(\omega_0 t/80)^2) \times \cos(\omega_0 t)$, where the excitation frequency $\omega_0 = 2\pi/T_0$ lies in the bulk gap. For the spin-up / down excitation, displacements of the six sites 1 to 6 is the same except for a phase difference 120° between adjacent one, and all the six sites rotates clockwise / counter-clockwise around equilibrium center. It is clearly seen that the spin-up excitation only induces clockwise elastic wave propagation along the interface, while the spin-down excitation transports oppositely (see Supplemental materials for animation). Despite the curved interface has some sharp corners, e.g., 80° and 110°, the wave packet can still flow perfectly along the interface without obvious backward reflection or scattering.

## 5. Conclusion

We demonstrate that three types of topological phase, mimicking the QVHE insulator, Chern insulator and QSHE insulator of quantum systems, respectively, can be observed for in-plane elastic wave in a single mechanical honeycomb made of masses and springs. Each topologically non-trivial phase is associated with specific symmetry breaking realized by mass and spring constant patterning or the external source.

By introducing mass difference to the two sites of the unit cell, the breaking of $C_{6v}$ to $C_3$ symmetry causes the lifting of Dirac degeneracy. The TR symmetry ensures that, the circular polarization with opposite chirality, which constitutes the elastic version of valley pseudo-spin, is featured at opposite valleys, therefore QVHE insulator is formed. Further, with the addition of Coriolis' effect which breaks the TR symmetry, the system can be tuned to switch between QVHE and Chern insulator depending on competition of breaking strength between TR and inversion symmetry. The synthesis of QVHE and Chern insulators can produce much richer topologically protected edge states which are not reported before. On the other hand, the QSHE is exemplified by patterning the spring stiffness to form a supercell preserving the $C_{6v}$ symmetry. Thus the twofold Dirac degeneracy at the $\Gamma$ point of Brillouin zone, as well as the eigenstates corresponding to the two irreducible representations can be utilized to identify pseudo spin-orbit coupling effect. Particularly, non-trivial QSHE phase is obtained when the inter-cell stiffness is higher than that intra-cell ones.

Throughout the formulation, the $k \cdot p$ perturbation method which expands the Hamiltonian in the neighbor of band degeneracy is employed to establish for each class of topological phase an asymptotic continuum model. The physics during phase transition and associated topological invariants, protected edge wave mode and its decaying profile are analytically clarified, and are further validated numerically by Bloch wave analysis of domain-wall strip and transient simulations of finite lattice sample. We mention here that, one can also incorporate mass difference and rotating platform in the QSHE insulator lattice, and further explore the interaction between the QVHE, Chern and QSHE insulators in a 3D parameter space. By breaking the TR symmetry, the quantum anomalous Hall effect with a spin-locked chiral interface state is most likely to be observed. The proposed discrete model will provide useful guideline to the design of feasible continuum configuration for robust wave mitigation in context of elastic wave.

## Acknowledgments

This work was supported by the National Natural Science Foundation of China (Grant Nos. 11372035, 11472044, 11632003, 11521062, BX20180040, 11802017), 111 Project (No. B16003). Discussion from Xiao Hu is also acknowledged.

## APPENDIX I *k · p* perturbation method for discrete system

This section presents a general method to derive *k · p* perturbation model for discrete systems, which could be the actual discrete model of mass and spring, or discretized from continuum model by numerical method such as finite element method. For discrete system, the dispersion relation in general is obtained from a matrix eigenvalue problem,

$$\left(-\omega^2\mathbf{M}+\mathbf{K}(\mathbf{k})\right)\mathbf{u}(\mathbf{k})=\mathbf{0} \tag{A1}$$

where, $\mathbf{M}$ and $\mathbf{K}(\mathbf{k})$ are respectively mass and stiffness matrices, $\mathbf{k}$ is the Bloch wave vector, $\mathbf{u}$ represents the discretized DOFs in a unit cell. The mass matrix $\mathbf{M}$ is assumed to be constant, which is the case in this paper. Suppose a $N$-folded degeneracy occurs at a specific wave vector $\mathbf{k}_0$ with frequency $\omega_0$. Degenerated eigenvectors $\mathbf{u}_i$ are normalized as follows,

$$\begin{cases}\left(-\omega_0^2\mathbf{M}+\mathbf{K}(\mathbf{k}_0)\right)\mathbf{u}_i(\mathbf{k}_0)=\mathbf{0}\\ \mathbf{u}_i^\dagger(\mathbf{k}_0)\mathbf{u}_j(\mathbf{k}_0)=\delta_{ij}\end{cases},\quad i,j=1,\ 2\ ...\ N \tag{A2}$$

Where indices $i, j$ are used to denote the degenerated eigenvectors, $\delta$ is the Kronecker symbol and the symbol † represents the conjugate transpose.

Now consider a new system perturbed from the original one Eq. (A1) by small geometry or material variation. The corresponding matrix eigenvalue problem is,

$$\left(-\omega^2\mathbf{M}'+\mathbf{K}'(\mathbf{k})\right)\mathbf{u}'(\mathbf{k})=\mathbf{0} \tag{A3}$$

In general, the new system is no longer degenerated at $\mathbf{k}_0$, and its eigenvectors at $\mathbf{k}_0$ will not be the same as in Eq. (A2). However, if the perturbation is small, the eigenvectors should be nearly the same as the original system Eq. (A2). Now adopting the traditional *k · p* perturbation method (Slonczewski and Weiss, 1958), eigenvectors of the perturbed system at $\mathbf{k}$ around $\mathbf{k}_0$ can be approximated as linear combinations of eigenvectors in Eq. (A2) of the original degenerated system. This technique significantly reduces the dimension of the eigenvalue problem, and transforms the eigenvalue problem Eq. (A3) into a new one with just $N$ DOFs. Follow this technique, eigenvector $\mathbf{u}'(\mathbf{k})$ for the new system is assumed,

$$\mathbf{u}'(\mathbf{k}) \approx c_i \mathbf{u}_i = c_1 \mathbf{u}_1 + c_2 \mathbf{u}_2 + \dots c_N \mathbf{u}_N \tag{A4}$$

Repeated indices should be understood as summation, and $c_i$ is unknown coefficient. Substituting the above expression into Eq. (A3) and multiplying both sides with the complex conjugate of $\mathbf{u}_j$, one could obtain $N$ equations in terms of the coefficient $c_i$,

$$\left(\mathbf{u}_j^\dagger (\omega^2 \mathbf{M}' - \mathbf{K}'(\mathbf{k}))\mathbf{u}_i\right) c_i = 0 \tag{A5}$$

This exactly solves the $N$ coefficient $c_i$. Further, expanding the stiffness matrix $\mathbf{K}'(\mathbf{k})$ at the wave vector $\mathbf{k}_0$ to the first order, one obtains a $k \cdot p$ effective model of the new system at $\mathbf{k}_0$ to the first order,

$$\mathbf{D}(\mathbf{k}_0, \omega)\psi = \mathbf{H}(\mathbf{k}_0, \Delta\mathbf{k})\psi, \quad \psi = \left(c_1, c_2, \dots c_N\right)^{\mathrm{T}}, \quad \psi^\dagger \psi = \mathbf{I} \tag{A6}$$

$$D_{ji}(\mathbf{k}_0, \omega) = \mathbf{u}_j^\dagger (\omega^2 \mathbf{M}' - \mathbf{K}'(\mathbf{k}_0))\mathbf{u}_i \tag{A7}$$

$$H_{ji}(\mathbf{k}_0, \Delta\mathbf{k}) = \left( \mathbf{u}_j^\dagger \frac{\partial \mathbf{K}'}{\partial \mathbf{k}} |_{\mathbf{k}_0} \mathbf{u}_i \right) \cdot \Delta\mathbf{k} \tag{A8}$$

Where $D_{ji}(\mathbf{k}_0, \omega)$ and $H_{ji}(\mathbf{k}_0, \Delta\mathbf{k})$ are the matrix elements of the second rank matrices $\mathbf{D}(\mathbf{k}_0, \omega)$ and $\mathbf{H}(\mathbf{k}_0, \Delta\mathbf{k})$, respectively. High order effective model for the perturbed system can be derived similarly by expanding the stiffness matrix $\mathbf{K}'(\mathbf{k})$ to higher order.

## APPENDIX II Derivation of the effective model

This section presents some detail to derive the effective model for the valley Hall insulator and spin Hall insulator. For the valley Hall phase, eigenstates for wave vectors close to the valley $\mathbf{K}'$ can be expressed as linear combination of the valley states $\mathbf{u}_p(\mathbf{K}')$ and $\mathbf{u}_q(\mathbf{K}')$. Follow the method in **APPENDIX I**, an effective model is derived,

$$\mathbf{D}(\mathbf{K}', \omega)\psi = \mathbf{H}(\mathbf{K}', \Delta\mathbf{k})\psi \tag{B1}$$

$$\mathbf{D}(\mathbf{K}', \omega) = \begin{pmatrix} m_p & 0 \\ 0 & m_q \end{pmatrix} \omega^2 - \frac{m_p + m_q}{2} \begin{pmatrix} 1 & 0 \\ 0 & 1 \end{pmatrix} \omega_0^2 \tag{B2}$$

$$\mathbf{H}(\mathbf{K}', \Delta\mathbf{k}) = \frac{\sqrt{3}at}{4} \begin{pmatrix} 0 & i\Delta k_x + \Delta k_y \\ -i\Delta k_x + \Delta k_y & 0 \end{pmatrix} \tag{B3}$$

where, $\Delta k_x$ and $\Delta k_y$ are incremental wave vector relative to $\mathbf{K}'$, $\omega_0 = (3t/(m_p + m_q))^{1/2}$ is the degenerate frequency of an imaginary system with equal mass $(m_p + m_q)/2$ for the sites $p$ and $q$. Considering an first order expansion of the eigen-frequency $\omega$ at $\omega_0$, we have

$$\mathbf{D}(\mathbf{K}',\omega) \approx \omega_0^2 \frac{m_p - m_q}{2}\begin{pmatrix} 1 & 0 \\ 0 & -1 \end{pmatrix} + 2\omega_0 \Delta\omega \frac{m_p + m_q}{2}\mathbf{I} \tag{B4}$$

In which, a second order small quantity $(m_p - m_q)\,\Delta\omega$ is omitted, and $\Delta\omega = \omega - \omega_0$ represents frequency deviation. After some simplification, an effective model Eq. (9) is derived for the valley $\mathbf{K}'$. Effective model for the other valley $\mathbf{K}$ can be derived similarly.

For the spin Hall insulator, eigenstates for wave vectors close to the Brillouin zone center $\mathbf{\Gamma}$ is approximated by the constructed pseudo spin basis $\{|p^+\rangle, |d^+\rangle, |p^-\rangle, |d^-\rangle\}$. An effective model is derived similarly following the above standard method,

$$\mathbf{D}(\mathbf{\Gamma},\omega)\psi = \mathbf{H}(\mathbf{\Gamma},\Delta\mathbf{k})\psi \tag{B5}$$

$$\mathbf{D}(\mathbf{\Gamma},\omega) = m_0(\omega^2 - \omega_0^2)\begin{pmatrix} 1 & 0 & 0 & 0 \\ 0 & 1 & 0 & 0 \\ 0 & 0 & 1 & 0 \\ 0 & 0 & 0 & 1 \end{pmatrix} + \frac{t_o - t_i}{2}\begin{pmatrix} 1 & 0 & 0 & 0 \\ 0 & -1 & 0 & 0 \\ 0 & 0 & 1 & 0 \\ 0 & 0 & 0 & -1 \end{pmatrix} \tag{B6}$$

$$\mathbf{H}(\mathbf{\Gamma},\Delta\mathbf{k}) = \frac{t_o}{4}\begin{pmatrix} \frac{1}{2}\Delta k^2 a^2 & (i\Delta k_x + \Delta k_y)a & 0 & 0 \\ 0 & -\frac{1}{2}\Delta k^2 a^2 & 0 & 0 \\ 0 & 0 & \frac{1}{2}\Delta k^2 a^2 & (i\Delta k_x - \Delta k_y)a \\ 0 & 0 & 0 & -\frac{1}{2}\Delta k^2 a^2 \end{pmatrix} \tag{B7}$$

where, $\Delta k_x$ and $\Delta k_y$ are also incremental wave vectors, $\Delta k^2 = (\Delta k_x)^2 + (\Delta k_y)^2$, and $\omega_0 = ((t_o + 2t_i)/m_0)^{1/2}$. Other off-diagonal terms in $\mathbf{H}$ are neglected here, since they only contribute as higher corrections to the frequency. The effective model Eq. (24) can be easily derived after some math operations.